\documentclass[acmlarge,nonacm]{acmart}

\usepackage{booktabs} 
\usepackage{soul}

\usepackage[ruled]{algorithm2e} 

\renewcommand{\hl}[1]{#1}

\acmJournal{IMWUT}
\acmVolume{0}
\acmNumber{0}
\acmArticle{0}
\acmYear{2021}
\acmMonth{1}

\setcopyright{acmcopyright}

\begin{document}
\title{AirWare: Utilizing Embedded Audio and Infrared Signals for In-Air Hand-Gesture Recognition}
\author{Nibhrat Lohia}
\authornote{The corresponding author}
\orcid{1234-5678-9012-3456}
\email{nlohia@smu.edu}
\author{Raunak Mundada}
\authornote{The first two authors are equal contributors to this work}
\email{rmundada@smu.edu}
\affiliation{%
  \institution{Southern Methodist University}
  \department{Lyle School of Engineering}
  \city{Dallas}
  \state{TX}
  \postcode{75205}
  \country{USA}
}
\author{Arya D. McCarthy}
\affiliation{
	\institution{Johns Hopkins University}
    \department{Computer Science}
    \city{Baltimore}
    \state{MD}
    \postcode{21218}
    \country{USA}
}
\authornote{Conducted at Southern Methodist University}
\author{Eric C. Larson}
\affiliation{%
  \institution{Southern Methodist University}
  \department{Lyle School of Engineering}
  \city{Dallas}
  \state{TX}
  \postcode{75205}
  \country{USA}
}

\begin{abstract}
We introduce AirWare, an in-air hand-gesture recognition system that uses the already embedded speaker and microphone in most electronic devices, together with embedded infrared proximity sensors. Gestures identified by AirWare are performed in the air above a touchscreen or a mobile phone. AirWare utilizes convolutional neural networks to classify a large vocabulary of hand gestures using multi-modal audio Doppler signatures and infrared (IR) sensor information. As opposed to other systems which use high frequency Doppler radars or depth cameras to uniquely identify in-air gestures, AirWare does not require any external sensors. In our analysis, we use openly available APIs to interface with the Samsung Galaxy S5 audio and proximity sensors for data collection. We find that AirWare is not reliable enough for a deployable interaction system when trying to classify a gesture set of 21 gestures, with an average true positive rate of only 50.5\% per gesture.
To improve performance, we train AirWare to identify subsets of the 21 gestures vocabulary based on possible usage scenarios. We find that AirWare can identify three gesture sets with average true positive rate greater than 80\%  using 4--7 gestures per set, which comprises a vocabulary of 16 unique in-air gestures.
\end{abstract}

%
%

\ccsdesc[500]{Human-centered computing~Gestural input}
\ccsdesc[500]{Human-centered computing~Ubiquitous and mobile devices}
\ccsdesc[500]{Computing methodologies~Machine learning}
\ccsdesc[500]{Computing methodologies~Neural networks}
%
%


\keywords{Doppler, Gesture Recognition, Deep Learning}

\maketitle

\section{Introduction}

Communicating through hand gestures is ubiquitous across cultures. 
Incorporating hand gestures into machine interaction has proven difficult because one must reliably detect the gestures and infer their meaning~\cite{lee2010search}.
Even so, with the increasing variety of devices that can interact with humans in more natural ways, in-air hand gesture recognition systems have grown in popularity.
Major technology companies like Google \cite{Soli}, Microsoft \cite{gupta2012soundwave}, Amazon, and HP have released devices that  recognize some basic in-air hand gestures. 
To achieve reliability, these devices employ specialized sensors or vision systems, like the Microsoft Kinect, which increase cost and reduce the potential ubiquity of the device.
This is worrying because the use of in-air hand gestures to interact with a machine is often desired only in niche scenarios when touch is inappropriate or difficult.
This is especially true for mobile devices (1) when the  device is small and touch is harder to use without occluding screen content (such as a watch) or  (2) in situational impairments, like wearing gloves or cooking, when hands get dirty and touching a smart-phone or tablet is not desired~\cite{dumas2009multimodal,wobbrock2006future}. There are also scenarios where in-air gestures may add to the user experience, such as gaming or productivity applications.

In this study, we investigate methods for detecting and classifying in-air gestures using commodity sensors on a smartphone.
More specifically, we present AirWare, a system that fuses the information from the on-board infrared (IR) proximity sensor of a Samsung Galaxy S5 with the Doppler shifts detected by the microphone.
Like previous work \cite{aumi2013doplink,chen2014airlink,gupta2012soundwave,sun2013spartacus} we play an inaudible tone from the speakers and record from the microphone continuously.  
Using signal processing and machine learning, we fuse the parameters of the IR proximity sensor with the Doppler features to predict a large vocabulary of different in-air gestures.

Our approach differs from previous work in that we (1) combine complementary sensors that are already embedded in the mobile phone and (2) attempt to classify a relatively large vocabulary of different gestures.
Most previous works only cover a few basic interactions (like panning), which does not provide a rich interaction modality.  

To inform the design and evaluate AirWare, we conducted a user study with 13 participants that performed each gesture several times (load balanced in terms of presentation order). We show that, on average, AirWare can recognize the full gesture vocabulary with only 50.47\% average true positive rate per gesture per user on 21 gestures. We conclude that the full 21 gesture vocabulary is not accurate enough to support user interaction.  However, using various reduced vocabularies of 4--7 gestures, AirWare can achieve average true positive rates greater than 75\% for each reduced vocabulary. In total, the reduced gesture sets comprise 16 unique gestures. We enumerate our contributions as follows:

\begin{enumerate}
\item We investigate the performance of fusing Doppler gesture sensing techniques with the embedded IR proximity sensor using various machine learning algorithms. We also compare the performance of a number of convolutional neural network architectures.
\item A human subjects evaluation: we validate the technology in a user study with 13 participants.
\item We compare two different methods for collecting gesture data. The first requires the IR sensor to be activated and the second is a free-form system. We conclude that the free-form system creates variability in the way gestures are performed such that the machine learning algorithms cannot readily identify the gestures. Therefore, the AirWare system requires users to be instructed on how to perform each gesture.
\item We investigate personalized calibration to boost the recognition true positive rate of the classifier, as well as providing an out-of-the-box gesture recognition system, showing that user calibration improves the performance of AirWare. We also investigate the amount of training data required to calibrate the AirWare system, concluding that 2-3 examples per gesture are needed to properly calibrate the system.
\item We investigate a number of reduced gesture vocabularies that tailor to different application use cases, showing average true positive rates greater than 80\% among subsets of gestures. When comprising the different subset, we conclude that AirWare can support about 16 total gestures. We also conclude that some gesture combinations cannot be supported, such as simultaneously identifying pans and flicks.
\end{enumerate}


\section{Related Work}

Dating back to 1980, in-air gesture sensing was achieved using commodity cameras with a high degree of success~\cite{rubine1991automatic}. The RGB image of a user-facing camera was used to detect and follow hand movements~\cite{hilliges2009interactions,rautaray2015vision}. Even so, privacy concerns and the requirements of processing video (battery life, lag time) limited the impact of the technology~\cite{hinckley2003synchronous,locken2012user,song2014air,suarez2012hand}. To mitigate these concerns, researchers have been innovating in how they sense hand motions. In Samsung's Galaxy S4 and S5 smart-phones, a dedicated infrared proximity sensor is used to detect hand motions above the phone, sensing velocity, angle, and relative distance of hand movements. The estimation is coarse, but allows for recognition of a number of panning gestures. We note that the IR proximity sensor is not unique the Samsung smartphones, but is used by a number of different manufacturers. However, these manufacturers typically do not provide access the the sensor via a developer API. With this in mind, the AirWare methodology could be applied to these phones in the future, once the sensors become accessible.

~\citet{gupta2012soundwave} used an inaudible tone played on speakers and sensed the Doppler reflections to determine when a user moved their hands toward or away from an interface. \citet{aumi2013doplink}, \citet{bannis2014adding}, \citet{sun2013spartacus}, and \citet{chen2014airlink} extended this work to detect pointing and flick gestures toward an array of objects (including smartphones) using the Doppler effect. These previous works typically recognize 2--4 gestures and many employ more than one set of speaker and microphone. In contrast, AirWare attempts to classify 21 gestures and various subsets ranging from 4--7 gestures per set. AirWare is able to classify such a large vocabulary of different gestures because the IR and audio Doppler combination provides complementary sensing information without any external sensors.

There have also been a number of innovative solutions that use infrared illuminating pendants~\cite{starner2000gesture}, magnetic sensors~\cite{chen2013utrack}, side mounted light sensors~\cite{butler2008sidesight}, and even the unmodified GSM antenna radiation~\cite{zhao2014sideswipe}. 
However, AirWare is more ambitious in the vocabulary size of gestures we attempt to classify, as well as unique in terms of the fused sensor outputs investigated. Moreover, AirWare does not use external sensors, but instead employs already embedded sensors from the mobile phone.

\citet{raj2012ultrasonic} review the HCI uses of Doppler sensing from ultrasonic range finders. Although this requires an extra sensor, it uses many of the same techniques as audio Doppler sensing. By sending a set of "pings" into the environment, the proximity of the hand (or any object) can be ascertained with high accuracy. AirWare shares some similarity in sensing techniques as we also employ a proximity sensor. Even so, AirWare uses the embedded Galaxy S5 proximity sensor, which is considerably less precise than the external ultrasonic sensor employed by previous works.

\citet{butler2008sidesight} produce IR sensor boards attached to a mobile device that succeed at identifying single- and multi-finger gestures adjacent to a device; however, this work does not address in-air gestures, use built-in hardware, combine modalities, or address differentiating a large gesture vocabulary.

\begin{figure}[t]
\centering
  \includegraphics[width=0.6\columnwidth]{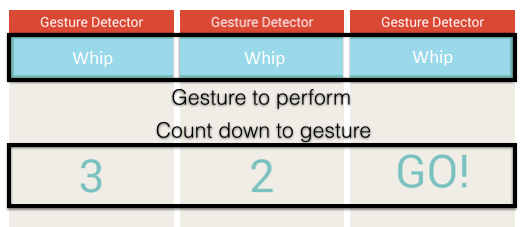}
  \caption{Progression of the AirWare interface used in our data collection.}
  \label{fig:Figure 1}
\end{figure}

\citet{kim2016hand} use micro-Doppler signatures with a convolution neural network. These Doppler signatures are measured by continuous wave Doppler radar at 5.8GHz (rather than the audio Doppler signal employed by AirWare). Their work encourages us to utilize convolutional neural networks to recognize in-air gestures. Their system classifies a set of 10 gestures with a five-fold cross validation accuracy of 85.6\%. For a reduced gesture set, the accuracy increases to 93.1\%. Although the system performs well, hand motions of the gestures are controlled. For example, swiping left to right was a quick snap that involved the wrist and all five fingers. However, for swiping right to left, the wrist was no longer stationary and moved with only three fingers involved. AirWare, however, does not impose such restrictions on the users motions and, instead, relies on the IR proximity sensor measurement along with Doppler signatures to classify the gestures. Moreover, AirWare employs convolutional networks on multiple sensor sources using embedded sensors, rather than an external, high frequency RF radar system.

Similar to \citeauthor{kim2016hand}, \citet{kim2016human} use deep convolution neural networks with continuous wave Doppler radars (i.e., using RF signals) for detection of humans and, to some degree, hand gestures. It is important to note that Kim uses specialized high frequency radar equipment. However, AirWare employs low-frequency audio Doppler from commodity hardware. Moreover, we employ infrared proximity sensors to create additional information that Doppler shifts do not capture, such as the occurrence of movements transverse to the sensing apparatus. 
\citet{raj2012ultrasonic} used a similar high-frequency device setup to classify most basic gestures. The results from these works were promising but fall in the same category of adding external sensors for recognition. Most have used ultra high frequency sonars for collecting Doppler signatures, with some specific spatial arrangements in some cases, thus causing greater frequency shifts which are relatively easier to classify.

\section{Theory of Operation}
In this section, we outline the different properties of each sensing modality: Doppler and IR proximity. We also posit an argument for why combining these modalities is inherently complementary. These sections also summarize the ways in which we access and pre-process each signal.

\begin{figure}[t]
  \includegraphics[width=0.6\linewidth]{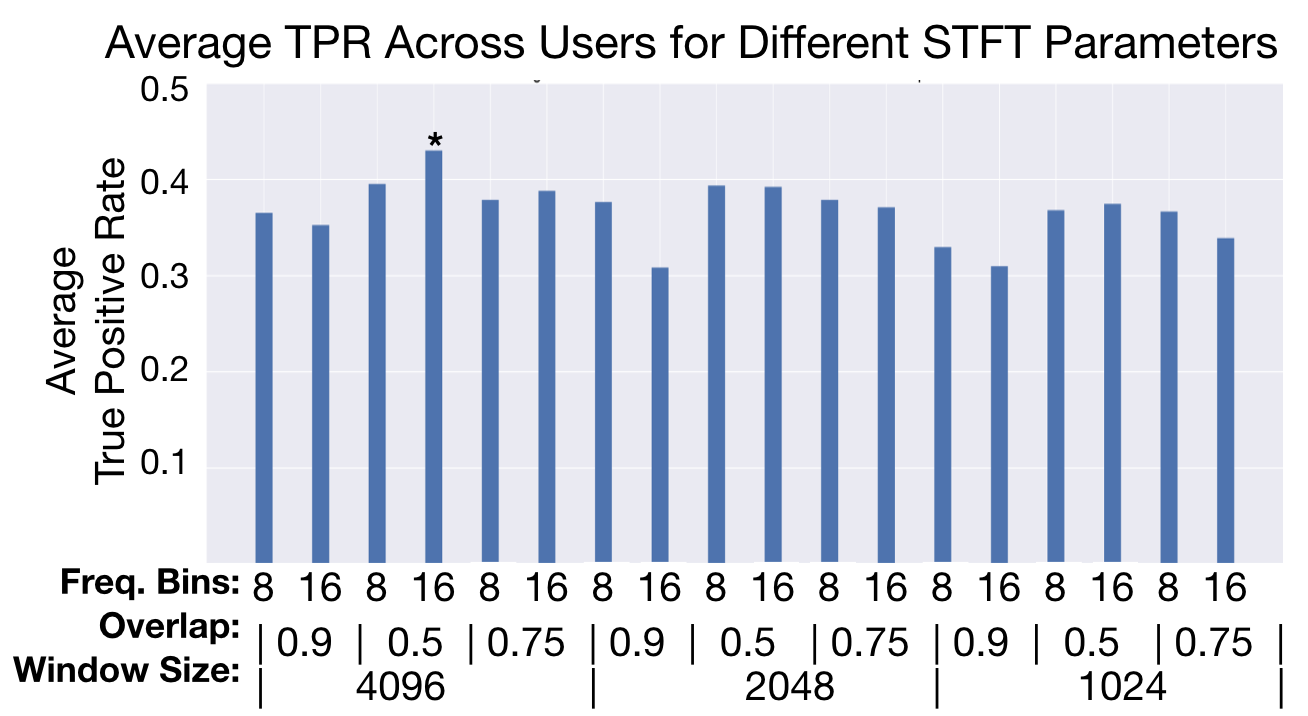}
  \caption{Average true positive rate for different short time Fourier transform parameters. A grid search reveals that a window size of 4096, 50\% overlap, and use of $\pm 16$ bins from $f_0$ performs the best.}
  \label{fig:STFT_Exp}
\end{figure}

\subsection{Audio Doppler from Speaker and Microphone}
Audio Doppler sensing for gesture recognition is discussed in detail in a number of papers~\cite{gupta2012soundwave, aumi2013doplink}.
Our method generally follows that of other papers. We play an 18~kHz sine wave from the speakers of the mobile phone, while continuously sampling from the microphone at 48~kHz. This means that a constant 18~kHz sine wave will be sampled from the microphone. When an object moves toward or away form a stationary phone, the microphone can detect Doppler frequency reflections. These manifest as additional reflections, added to the 18~kHz sine wave. The change in frequency is given by:
\begin{equation}
\Delta f = \frac{f_0v}{c}\cos(\theta)
\end{equation}
where $c$~is the speed of sound, $v$~is the velocity of the object, $\theta$ \hl{is the angle between the motion of the object and the microphone}, and $f_0$~is the frequency of the sine wave played from the speakers. 
When analyzing the signal, faster motions toward the microphone result in more pronounced \hl{frequency increases. Movement away from the microphone results in frequency decreases. The angle between the object movement and the microphone is a key factor. Movement transverse to the microphone results in no frequency shift. Movement directly toward or away the microphone maximizes the possible frequency change. As such, the frequency reflections form different hand gestures at different angles will manifest differently in the Doppler signal. More than just frequency, the surface area of the object determines the magnitude changes of the frequency reflections. This means it is possible to detect the difference between waving a finger at the microphone versus waving a hand at the same velocity.}
To capture these changes in the frequency over time, like previous approaches, we use the short time Fourier transform (STFT).
Specifically, we use a sampling rate of 48~kHz.  
We use a Hamming window to reduce spectral leakage. \hl{Other parameters of the STFT need to be chosen to trade off resolution in time and resolution in frequency (i.e., a classic signal processing problem).} It was unclear what trade-off between time and frequency to employ \hl{in order to capture the motion of the hand and fingers. For instance, choosing a large window size would increase our frequency resolution (i.e., our ability to discern Doppler shifts), but would also reduce our time resolution (i.e., our ability to observe quick movements). As such, we decided to grid search different STFT parameters. Each different configuration resulted in slightly different frequency profiles that could be used as features in our machine learning algorithms. We investigated 18 different configurations based upon the following combinations of parameters:}

\begin{itemize}
\item window size (and FFT size) of 1024, 2048, and 4096 samples
\item overlap between windows of 25\%, 50\%, and 75\%
\item \hl{number of frequency bins above and below $f_0$ to include as features, 8 or 16}
\end{itemize}

\begin{figure}[t]
  \includegraphics[width=\linewidth]{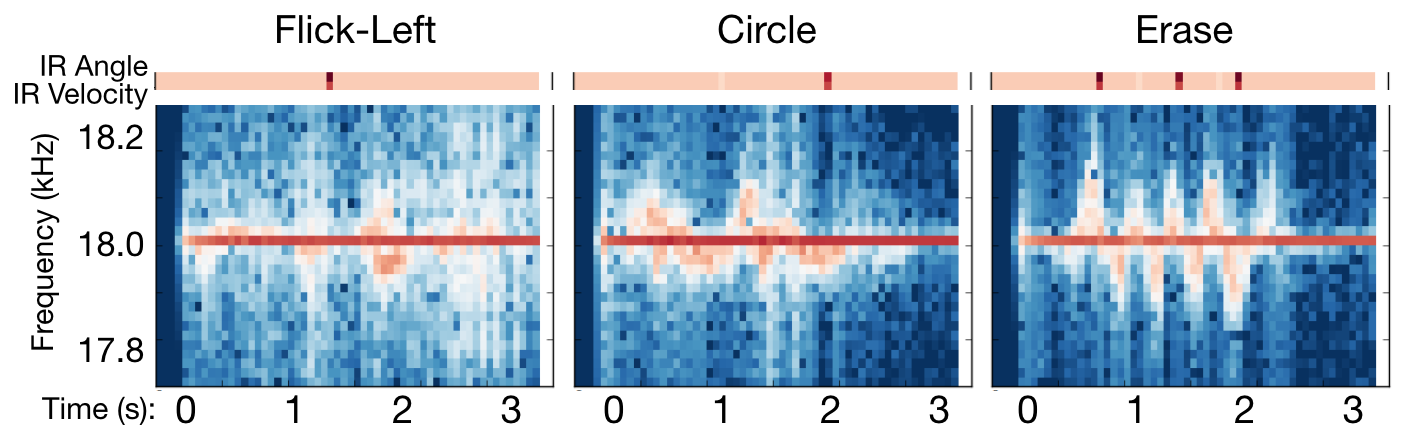}
  \caption{Zoomed spectrogram of three different gestures. Top of plots show the output of the IR sensor angle and velocity.}
  \label{fig:spec}
\end{figure}

\hl{The average true positive rate per gesture per user of the different configurations is shown in} \autoref{fig:STFT_Exp}.\hl{ Many configurations result in similar performance, but the best configuration was found to be: window size of 4096 points, 50\% overlap, and 16 bins above and below $f_0$. More details about the machine learning and cross validation techniques are discussed later.} We save the STFT for three seconds of time data (discarding the initial startup windows). An example of the STFT \hl{with the best found configuration} can be seen in \autoref{fig:spec}.

To normalize and control dynamic range, we take the decibel magnitude of the STFT. \hl{The implementation of the STFT grid search and feature extraction techniques have been made open source and are available at} \cite{opensourceRaunak}.

\subsection{IR Sensing}
The infrared proximity sensor for the Samsung Galaxy S5 is a set of four infrared sensors surrounding an infrared LED. Infrared light reflects back towards the four sensors when an object, like the hand, is above any of the sensors. By detecting which sensors are activated first and in which order, the sensor can infer what angle an object is moving laterally. The time difference in which each of the sensors is excited determines the approximate velocity of the object as it enters or exits the sensor area. When coupled with Doppler sensing, these two sensing modalities have a number of complementary features.  

\subsection{Complementary Sensors}
While Doppler sensing can provide rich information about the direction of motion towards or away from the microphone, about the relative velocity of movement, and \hl{the relative surface area of the object}, it is blind to absolute trajectory. That is, movement right and left  can look identical because they lie in the same plane. \hl{Moreover, perpendicular movement to the microphone may not cause any Doppler shifts and different combinations of velocity and incidence angle can manifest similarly in the Doppler signal.} For a number of in-air gestures, this directionality and velocity are critical to understanding the gesture. Fortunately, these lateral movements are exactly what the IR sensor is designed to detect, \hl{which adds a complementary source of information}. However, the use of an IR proximity sensor is not a panacea. The infrared sensor is often blind to gestures that occur at a distance from the sensor and it is often unable to distinguish motions that have a ``straight-on'' trajectory. In this way, the sensing modalities of infrared and Doppler are quite complementary as outlined below:

  \textbf{Doppler Sensing:} 
  \begin{itemize}
  \item Sensitive to motion towards and away from the microphone but agnostic to lateral motions. \hl{Angle of motion results in different reflected frequencies.}
  \item Sensitive to the surface area of the object. \hl{Larger objects result in larger magnitude reflections.}
  \item Sensitive to the overall velocity towards or away from the microphone. \hl{Higher velocity movements result in different reflected frequencies}
  \item Sensitive to motions at both far and near distances from phone.
  \end{itemize}
  
  \textbf{IR Proximity Sensor:}
  \begin{itemize}
  \item Sensitive to motion lateral to sensor and, to some degree, motions toward the sensor
  \item Can discern angle of lateral motions but not when motion is directed towards sensor 
  \item sensitive to the overall velocity of lateral motions
  \item Only sensitive to motions that occur relatively close to the sensor
  \end{itemize}

Given these properties, it is easier to understand why the combination of sensors reveals complementary details about in-air hand motions. While the IR sensor can detect panning motions, it cannot distinguish if the motion comes from fingers or the palm of the hand. While the Doppler signal shows velocity of the hand as it passes by the microphone, it cannot discern if the motion is from left to right or from top to bottom and vice-versa. These types of differences are imperative to understand a large vocabulary of in-air hand gestures. We investigate several methods to combine the IR sensor stream and the STFT through traditional machine learning algorithms and convolutional neural networks.

\section{Spectrogram Processing}
In this stage, our aim is to process and combine the features from the STFT and the IR sensors so that they can be analyzed by a machine learning algorithm.
We start by finding the magnitude of the generated 18~kHz tone across the entire STFT,~$ M_{0 \mathrm{dB}}(t) $, where~$ t $ denotes the frame number~$(0, 1,\ldots,99)$. We then isolate a band of magnitudes around the frequency in a range of frequency bins above and below~$f_0$. \hl{The number of bins above and below the~$f_0$ is a grid-searched parameter in our analysis. We found that using 16 bins above and below the~$f_0$ is sufficient for classification.} These ranges are shown in \autoref{fig:spec}.
During this feature extraction, we also eliminate the magnitude of~$f_0$ from the spectrogram, as the value is relatively constant in magnitude and therefore has limited predictive capability as a feature. 

After processing the STFT, we process the features extracted from the Samsung Galaxy S5's IR sensor. The sensor interface uses a ``push'' style API where the application subscribes to notifications when the sensor is activated. The notification includes the speed (a normalized value between 0 and 100) and angle of any detected movements. The angle is an average of the entering and exit angles. However, gestures that do not move laterally across the sensor typically register as having zero velocity and zero degree angle because the sensor cannot validly estimate the movement (but detects that an object is close to the sensor). Each time we are notified of a movement, we log the event and time-stamp when the event occurred. 

\subsection{Segmentation}
\hl{When a user performs a gesture, it may or may not activate the IR sensor. We performed two rounds of data collection. The first did not require that users activate the IR sensor with the gesture and the second did require that the IR sensor be activated. The segmentation procedure differs slightly between these two scenarios. When we required the user to activate the IR sensor, segmentation was straightforward: we buffer the audio signal 1.25 seconds before and after the IR activation. When we did not require the IR sensor to activate, we buffered 1.25 seconds before and after any ``event of interest.''  We define this event to be when either the IR sensor is activated or when the magnitude of frequency bins directly greater than and less than $f_0$ increase by 10~dB. Intuitively, this occurs when there is enough motion to cause reflections of the Doppler audio signal. We also note that, when not requiring the IR sensor to be activated, we expect an increased number of false positives because any motion might trigger the segmentation algorithm. Positively, requiring the IR sensor to be activated by the gesture can be considered an effective means of reducing false positives. Negatively, it also requires users to manipulate their gestures in a way that they always trigger the sensor at the top of the phone. This limitation is discussed  in more depth in the next section.}

Once all data is collected for all users, we employ normalization of each of the IR features (angle and velocity) and of the entire spectrogram magnitudes such that the all features are zero mean and unit standard deviation.  

\section{Experimental Methodology}
In our pilot tests, we asked participants to perform each of the 21 gestures  in the way ``that made the most sense to them.''  
In this way, we sought to collect more realistic data where participants could be trained simply from a textual prompt of what the gesture was, without explicit training or demonstration. Therefore, we thought the gesture would be more intuitive to the user (since they exhibit their internalization of the gesture, rather than mimicking a gesture they were shown). However, this data was never classifiable at a rate more than chance. We abandoned the idea that a large gesture vocabulary could be collected without explicitly demonstrating the gestures to participants. \hl{Based on our experience in the pilot study, we decided to update our methodology to include showing videos of the gestures being performed. Participants were then asked to perform the gesture to demonstrate their understanding.}  Therefore, all participants were shown how to perform the gestures and participants demonstrated their understanding to the researchers before data collection started. \hl{Practically, this also means that new users of AirWare would also need to go through the same instructional videos to learn how to perform the gestures in the vocabulary. We see this as a necessary limitation of the AirWare system: without an instructional phase, there is too much variability among the gestures performed to detect them reliably. }

\hl{Because our pilot study had uncovered that gesture consistency might be problematic}, we conducted a user study in two phases. We chose two phases because it was unclear how to define what a ``proper'' gesture consisted of. Each phase differed in what the data collection application judged to be a properly performed gesture. In the first phase, we collected gesture data from the participants for every gesture in our vocabulary regardless of whether the IR sensor was activated. That is, the user performed a gesture based upon their memory of how the gesture was performed from the instructional videos. In the second phase, we only informed the participant that a gesture was performed successfully when the IR proximity sensor was activated. \hl{That is, they were asked to repeat the gesture until they learned how to perform the gesture while also activating the sensor at the top of the phone screen.} In this way, users needed to manipulate the way they performed the gesture such that they understood where the proximity sensors was physically located on the phone and how to activate it with each gesture.   

Different users participated in each phase to protect against crossover effects. That is, no user participated in both phases of the data collection. \hl{We show later on that requiring the IR sensor to be activated greatly increases the ability of the machine learning algorithm to correctly identify the gesture. Practically, this means that the AirWare system will almost certainly require an ``instructional application'' that trains users to perform the gestures, and then verifies that the user understands how to perform the gesture such that the IR sensor is activated. While this is an additional limitation of the system because it imposes constraints on the gestures, such an instructional application would likely be required no matter what, as learning to perform 21 gestures for any user  without some instruction can be considered a daunting task.}

In the first phase, 8 participants were recruited from university classes (age range: 19-30, Male: ~60\%). During a session, participants were introduced to the AirWare data collection mobile application and a demonstration of all the gestures were given via the video recording. Participants were then instructed to show the researcher each gesture. They weren't told about the sensor locations on the phone. The ambient environment was relatively quiet and without many acoustic disturbances. Participants were instructed to hold the smart-phone in one hand and perform gestures ``above'' the phone with the other hand. 

In the second phase, \hl{13} participants were recruited (\hl{age range: 19-30, Male: ~66\%}). Participants were similarly introduced to the data collection application but were also instructed about the location of IR proximity sensor on the phone (as described). The user interface showed the participant whenever the IR sensor detected a movement through an animated label on the application. The gesture data was registered only when the IR sensor detected a movement; otherwise, the interface prompted the user to repeat the gesture. On average, users  had some initial trouble learning how to manipulate the sensor for some gestures such as ``tap'' but were quickly able to alter their strategy to tap towards the top of the phone (where the IR sensor was located).  \hl{All users were able to successfully activate the IR sensor after two or three trials per gesture.}

\begin{figure}[t]
  \includegraphics[width=\linewidth]{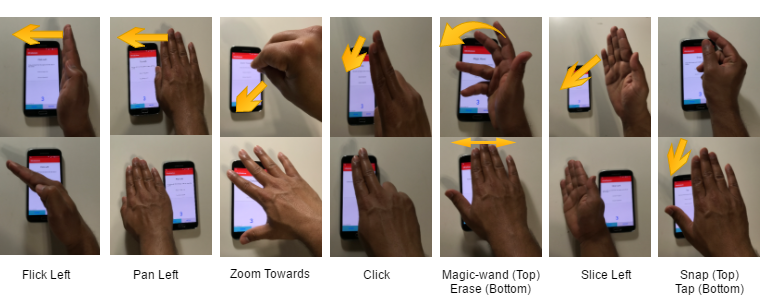}
  \caption{Examples of the different gestures predicted in the AirWare vocabulary.}
  \label{fig:gestures}
\end{figure}

\subsection{Gesture Vocabulary}
Participants performed 21 different gestures as instructed on the screen of the phone via a custom data collection app. A gesture name would appear on the screen and the user would perform the in-air gesture (\autoref{fig:gestures}). All sensor data was saved locally on the phone for later processing. Users went through each gesture one time as practice (practice data was not used in analysis) and then were presented with a random permutation of the gestures. For participants in the second phase, the practice session lasted as long as was needed for the subject to learn how to activate the IR sensor. In all, each participant performed between 5 and 10 iterations of each gesture. The different number of gestures per participant is an artifact of the way the gestures were randomly presented. We let participants perform gestures for 45~minutes and then ended the session. On average, each participant performed about 250~gestures. \hl{Note that the smartphone was used for data collection only. Subsequent analysis was performed offline.}

The initial gestures chosen were based upon an informal review of other in-air gesture sensing systems. After the initial pilot study, we refined the gestures based upon informal discussions about what gestures were most intuitive to perform. The final gesture set consisted of: 

\begin{itemize}
\item Flick left/right/up/down (quick hand movement from wrist)
\item Pan left/right/up/down (hand flat, movement from elbow)
\item Slice left/right (a fast ``sword'' motion diagonally across phone)
\item Zoom in/out (whole hand)
\item Whip (motion towards phone like a holding a whip)
\item Snap (similar to whip but snapping while moving toward the phone)
\item Magic wand (slow waving of the fingers towards the palm) 
\item Click, double click (with finger, but not touching screen)
\item Tap, double tap (with full hand)
\item Circle (circular motion above the phone, hand flat)
\item Erase (moving hand back and forth)
\end{itemize}

\hl{In our pilot studies, more gesture were included in the vocabulary. However, some gestures which users felt were awkward or unintuitive were removed. These included gestures such as hand wobble, finger wave in/out, and push/pull gestures.}


\section{Machine Learning Description}
To create, train and validate machine learning algorithms we use a combination of packages in Python. Specifically, we use the ``scikit-learn'' library~\cite{pedregosa2011scikit} and Keras~\cite{chollet2015keras} with the  TensorFlow~\cite{tensorflow2015-whitepaper} back-end. \hl{We chose to investigate several different machine learning baselines and also several different convolutional neural network architectures. It was unclear what neural network architecture and parameters of the architecture would be optimal, so we chose to train several variants and perform hyper parameter tuning for each architecture.}

\subsection{Baseline Models - Traditional ML algorithms}
\hl{For baseline comparison, we investigated several traditional machine learning algorithms including multi-layer perceptrons, linear support vector machines, and random forests. The doppler information is pre-processed using Principal Component Analysis (PCA) to reduce the dimensions. The IR information is averaged across time steps for each gesture. For each model, the hyper-parameters as well as the number of principal components for doppler information are selected via a randomized grid searching strategy. Based on our grid searching results, hyper-parameters for each model are as follows}:

 \textbf{Random Forest:}
	\begin{itemize}
		\item Number of trees/estimators: 1000
		\item Bootstrap: True
		\item Node split criterion: Gini-index
		\item Maximum Features: $\sqrt{N}$ 
		\item Number of Principal Components: 100
	\end{itemize}
    
 \textbf{Support Vector Machines:}
	\begin{itemize}
		\item Kernel: Linear kernel
		\item Penalty parameter: 10
		\item Number of Principal Components: 100
	\end{itemize}
    
 \textbf{Multi-layer Perceptron:}
	\begin{itemize}
    	\item Hidden Layer 1 unit size: 500
        \item Hidden Layer 2 unit size: 250
        \item Early Stopping: True
        \item Gradient Solver: Stochastic Gradient Descent
        \item Activation Unit: Tanh
        \item L2 Regularization: 0.01
        \item Number of Principal Components: 100
	\end{itemize}

We choose these algorithms because they span a wide variation of properties including various decision boundary capabilities such as linear versus arbitrary.

\subsection{Convolutional Network Architectures}
\begin{figure*}
  \includegraphics[width=\linewidth]{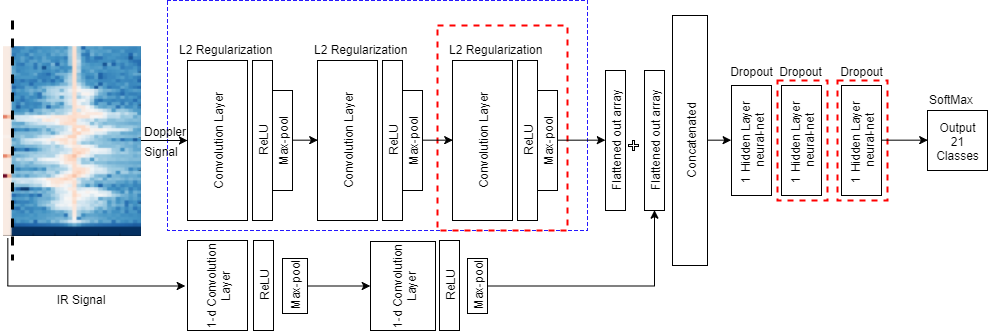}
  \caption{Diagram of the convolutional neural networks investigated. We vary the number of layers in the convolutional layers and number of dense layers as part of our analysis.}
  \label{fig:convnet}
\end{figure*}

\hl{In addition to these baseline machine learning models, we chose to investigate four different convolutional neural network architectures. The differences between each model come from the number of convolutional and dense layers employed, as well as the type of convolution employed (one dimensional versus two dimensional). The first three models all employed one-dimensional convolutions. This makes the processing more similar to approaches used in natural language processing than in image processing. In text processing, one-dimensional convolutional filters are  typically convolved with the word embedding matrix over a sequence}~\cite{severyn2015learning}. \hl{The fourth model employed two dimensional convolutional filters on the input spectrogram, which is more common in image processing. 

Each architecture follows from the basic diagram shown in} \autoref{fig:convnet}. \hl{In this architecture, the spectrogram and IR signals are processed separately, through similar convolutional branches of the network. They are then concatenated and passed to dense hidden layers.  The differences among the three models that employed one-dimensional convolutions is the depth of the convolutional and dense layers employed. The most simple architecture employs two convolutional layers and two dense layers, followed by an output layer. Another model employs three convolutional layers, and another model employs three dense layers. In all models, every convolutional layer is followed by a max pool layer. } $L_2$~regularization is used in all convolutional layers to minimize over-fitting. Rectified Linear Unit (ReLU) activations are used everywhere except the final layer to speed up the training and avoid unstable gradients. Finally a Softmax layer is used at the output which finally classifies the gestures into 21~different classes. \hl{To more clearly reference each of the four models, we refer to each model by number. Note that all models use two convolutional layers to analyze the IR proximity sensor, but different number of layers for the spectrogram branch of the network:}
\begin{itemize}
\item \hl{Model 1: 1D convolutional filters, two convolutional spectrogram layers, two dense layers}
\item \hl{Model 2: 1D convolutional filters, two convolutional spectrogram layers, four dense layers}
\item \hl{Model 3: 1D convolutional filters, three convolutional spectrogram layers, four dense layers}
\item \hl{Model 4: 2D convolutional filters, two convolutional spectrogram layers, two dense layers}
\end{itemize}



When training the network, we apply random perturbations to the input spectrogram and IR sequences to help avoid over fitting and increase generalization performance (i.e., data expansion). We randomly shift the data temporally up to~10\%. \hl{That is, we shift the entire spectrogram sequence forward or backward in time randomly by up to 10\%. The sequences are 2.5 seconds in duration, so this means that the the spectrogram and/or the IR stream might shift by 250~ms. } This is applied to the IR data and the spectrogram separately (resulting in different random time shifts). This helps with generalization performance because the Samsung Gesture API is somewhat inconsistent in the timings for when it provides the push notification that the IR sensor has been activated. Therefore this data expansion mirrors the actual use case well. 

\subsection{Hyper-parameter Tuning}
\hl{Based on the works of Bengio}~\cite{bengio2012practical} \hl{and Bergstra et al.}~\cite{bergstra2011algorithms} \hl{we chose to use the Tree-structured Parzen Estimator approach for hyper parameter tuning. These works established this estimation approach to be superior for tuning hyper-parameters compared to randomized search. During tuning, we only vary the number of convolutional filters and kernel size for the spectrogram branch of the network because the signal size is relatively more complex than the IR sensor stream. The filters applied to the IR signal are held constant at 2 one dimensional filters with kernel length of 2. The following parameters were tuned:}
\begin{itemize} 
\item \hl{L2 Regularization: Normal Distribution with mean 0.001 and s.d 0.0001}
\item \hl{Learning Rate ($10^{-X}$): [-6, -5, -4, -3, -2, -1, 0]}
\item \hl{Number of convolution filters: [8, 16, 32, 64]  }
\item \hl{Kernel Size: [2, 3, 5] }
\item \hl{Dropout: Uniform Distribution [0, 0.99]}
\item \hl{Number of hidden layer units: [32, 64, 128, 256, 512]}
\item \hl{Kernel weight initializers: [He normal and uniform distributions], [Glorot normal and uniform distributions], [LeCun normal and uniform distributions]}
\end{itemize}
\hl{Each of the four models underwent hyper parameter tuning. When comparing the different architectures, we use the best set of hyper parameters found for each architecture.}


\section{Results and Summary}
We divide our results into three overarching sections: comparisons between segmentations that require IR activation versus not requiring IR activation, classification with the full gesture set, and classification with multiple subsets of the gesture vocabulary. 

\subsection{IR Activation Segmentation Comparison}
In this section, we compare the predictive ability of gestures collected requiring that the IR sensor be activated versus not requiring the IR sensor to activate to segment gestures.  
Recall that these data sets are collected using separate experiments and different users. In each scenario, we train the models using leave-one-subject-out cross validation. That is, no subject's data is simultaneously used for training and testing.  For our evaluation metric, we choose the average true positive rate per gesture. Because class imbalance exists, accuracy is not a good indicator of performance as classes that occur less often will receive less weight in the evaluation. Moreover, binary scores like recall and precision are harder to  interpret when micro or macro averaged. Per-class true positive rate, alternatively, captures how well we perform for each gesture. \hl{For this analysis, we choose to use the random forest baseline model, as it is the best performing baseline model (discussed later).} \autoref{tab:irnoir} describes the per-class true positive rate for requiring versus not requiring the IR sensor to be activated \hl{for the random forest model}. The main conclusion we draw from \autoref{tab:irnoir} is that requiring the IR sensor to activate does increase the performance of the AirWare algorithm. Moreover, there are other advantages for requiring that the sensor be activated such as reducing false positives and reducing needless computation. \hl{This is because the audio Doppler signal is likely to result in a number of false positives from movement by the user and near the user; whereas the IR sensor is relatively robust to these types of noise. However, requiring that the IR sensor be activated also requires users to manipulate the way they perform in-air gestures to activate the sensor. In this way, the AirWare system will likely require some instruction to the users for how to reliably perform different gestures}. Thus, in the remainder of our analysis we only use the gesture set that requires IR activation to segment gestures. 

\begin{table}[t]
\caption{Average true positive rate per class for differing IR sensor activation.}
\centering
\begin{tabular}{l|l|l}
                     & \multicolumn{2}{l}{Per Class True Positive Rate} \\
IR Sensor Activation & Average                & STD Error               \\
\hline
\textbf{Required}, N=13            & \hl{38.92}\%                & \hl{0.01}                    \\
Not Required, N=8         & \hl{13.71}\%                & \hl{0.02}                    \\

Majority             & 5.34\%                 & N/A                    \\
Chance             & 4.76\%                 & N/A                    \\
\end{tabular}

\label{tab:irnoir}
\end{table}

\subsection{IR Activation and Doppler Signatures}

\begin{figure}[t]
\centering
  \includegraphics[width=0.5\linewidth]{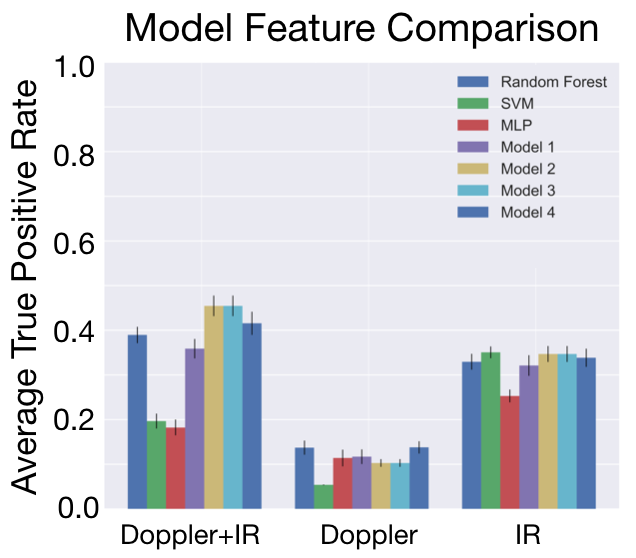}
  \caption{Average true positive rate per gesture per user comparison for using only IR activation, \hl{ only Doppler signatures and combined sensor information to classify the 21 class gesture set.}}
  \label{fig:IR_Doppler}
\end{figure}

\hl{In this section, we analyze the performance of using IR sensor data only, Doppler data only, and the combined sensors. With this analysis we seek to understand how advantageous it is to combine the sensors. To investigate this research question, } our network is modified to use only the IR activation information or only the Doppler signature information. From \autoref{fig:convnet} \hl{this corresponds to only using one of the input branches in the network. The performance of the model trained using individual sensor information is compared with the performance of the model trained using the combined sensor information.} The cross validation strategy used in \hl{all} cases is leave-one-subject-out, wherein we train the model on $N-1$ users' data and test the model on the $N$\textsuperscript{th} user data. \hl{If we only look at the best performing models from} \autoref{fig:IR_Doppler}, we see that we are able to achieve an average true positive rate of \hl{~35\%} per class with a standard error of~0.02\% when only the IR sensor information is used. \hl{Note that model 2 and 3 are identical when only using the IR signal branch.} \hl{In comparison, using audio Doppler only sensor information results in a best performing model with average true positive rate of ~13\% with a standard error of~0.03\%.} From \autoref{fig:IR_Doppler}, \hl{we can see that the performance with only IR information is better than performance of using Doppler only regardless of the machine learning model employed. However, when we combine the information from the two modalities, the performance improves for all convolutional neural network models and the random forest model. Thus, we conclude that combining the two sensing modalities is advantageous for in-air gesture recognition, resulting in a performance increase of about 10\% average true positive rate per gesture. The improvements are statistically significant based upon a two-tailed T-test ($p<0.01$). In all analyses in the remainder of the paper, we use the combined Doppler and IR sensor modalities as features for the machine learning models.}

\begin{figure}[b]
  \includegraphics[width=0.5\linewidth]{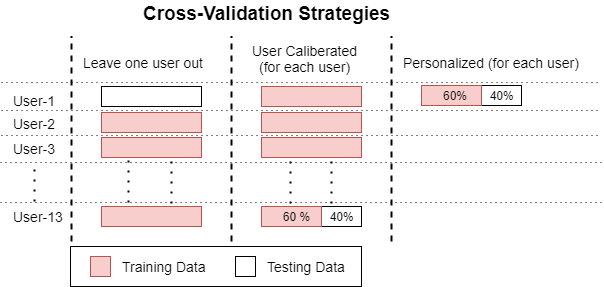}
  \caption{Overview of three different cross-validation strategies invetigated. }
  \label{fig:cvexplained}
\end{figure}

\subsection{Full 21-Gesture Vocabulary}
\hl{We now investigate the performance of the baseline machine learning models as well as different convolutional neural network architectures described in the previous section using different cross-validation strategies.}  We analyze the performance of the classifiers through the following cross validation strategies, each with its own practical implications. \hl{Also, an overview of each cross validation strategy is shown in} \autoref{fig:cvexplained}.

\begin{figure}[t]
\centering
  \includegraphics[width=0.5\linewidth]{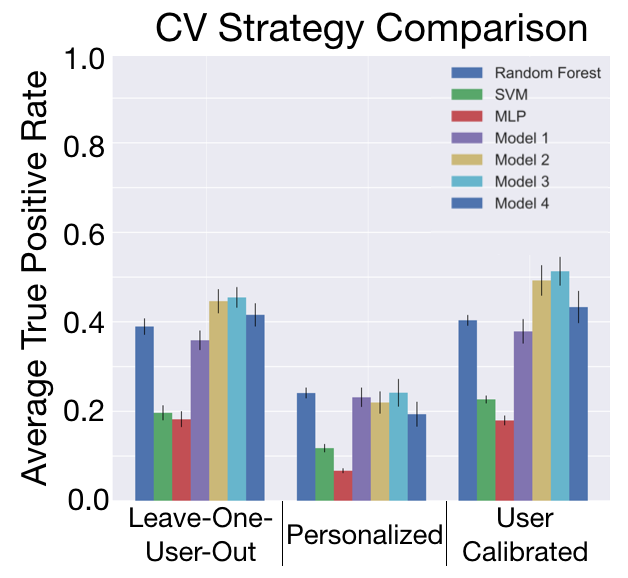}
  \caption{Average true positive rate per gesture per user comparison after combining the IR activation and Doppler signatures \hl{for all baseline and convolutional neural network models}}
  \label{fig:fullgestures}
\end{figure}

\subsubsection{Leave One User Out}
We explore the performance of our \hl{classifiers} using `leave one subject out' cross validation strategy. The strategy used in this case is to train the model on data from $N-1$ users and test it on the $N$\textsuperscript{th} user, \hl{as described in} \autoref{fig:cvexplained}. This approach analyzes whether we can classify the gestures successfully without \hl{requiring the system to be calibrated}. This implies that for practical implementations, we can directly use a pre-trained, out-of-the-box classifier to classify the gestures.  Through this strategy, we try to generalize the learning of our classifier across different types of users. \hl{This is the ideal scenario for a gesture system, requiring no user input or calibration before use.}
\hl{ As can been seen from } \autoref{fig:fullgestures}, \hl{average true positive rate per gesture per user ranges from approximately ~19\% to ~46\% for different classifiers. Our best performing model in this case is 'Model 3' which is a deeper network in terms of the number of convolutional layers and dense layers}. 

\hl{Breaking down the performance of Model 3 by users} \autoref{fig:bestmodel} \hl{we see that our network performs well, except for users 1, 3, and 4. We are able to achieve an average true positive rate per gesture per user of 45.19\% with a standard error of ~0.02\%. We conclude that the performance of the leave-one-user-out model is not sufficient for a gesture recognition system. Therefore, we explored other cross validation strategies that assume a calibration phase is employed.}

\subsubsection{Personalized Model for Each User}
\begin{figure}[t]
\centering
  \includegraphics[width=0.5\linewidth]{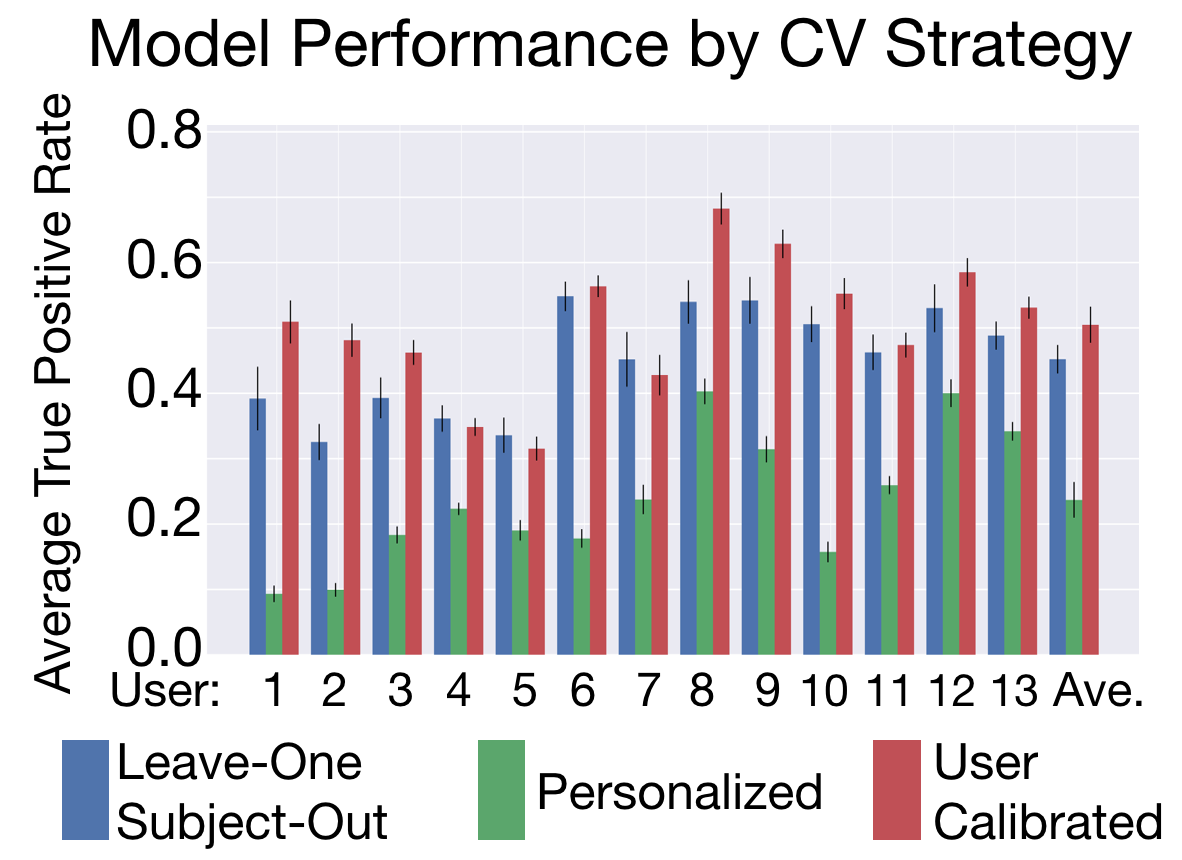}
  \caption{\hl{Average true positive rate per gesture per user comparison for the best performing architecture, Model 3. Results are shown for all employed cross validation strategies. }}
  \label{fig:bestmodel}
\end{figure}

In this analysis, we analyze the performance of our classifier by calibrating the model to each user. For each user, we perform a 5-iteration 60\% training and 40\% testing stratified, shuffled split, \hl{as described in} \autoref{fig:cvexplained}. The test size is chosen to make sure there are at least 2 instances of each class are present in the test data~\cite{pedregosa2011scikit}. In effect, we train \hl{13}~independent models using only data collected from a specific individual for training and testing. We would like to see if the variation in gesture performance within the user is able to predict the classes successfully. \hl{This cross validation mirrors a use case where users would need to provide example gestures to the system as a calibration phase. This is less ideal in terms of practical usage but may be necessary to increase performance.} \hl{From} \autoref{fig:fullgestures}, \hl{we see that, on average, the performance deteriorates for all the models as compared to leave one subject out. In this case, Random Forest performs equally well compared to 'Model 3' at approximately ~24\% average true positive rate per gesture per user. }
From  \autoref{fig:bestmodel}, \hl{we can see that none of the users benefit from a fully personalized model when compared with `leave one subject out' performance. We are able to achieve an average true positive rate  of~23.68\% per class with a standard error of~0.03\% across users for the 21-class gesture set.} \hl{It is unclear, however, if the personalized models do not perform consistently because the training data is limited. Convolutional networks tend to require large amounts of training data to perform well, so it is possible that the decrease in performance is due to a significantly smaller training set.} This motivates us to combine our two cross validation strategies in order to increase the amount of training data, but also employ a personalized calibration procedure.

\subsubsection{User Calibrated Model}
In this cross validation strategy, we combine knowledge from the previous two strategies to test the performance of the model. We first split the data based on the user; $N-1$ users' data in the training set. From the $N^{th}$ user data, we perform a 5-fold 60\%-40\% stratified shuffle split, as done for the personalized model. We then combine the training data from the $N-1$ users with the 60\% split of training data from the $N^{th}$ user and use the remaining 40\% of data from the $N^{th}$ user as a testing set, as shown in \autoref{fig:cvexplained}. The model performance for each user improves significantly for all users with this training strategy. Thus, the model learns from other users as well as the test user to classify the gestures of the test user. \hl{Note that this training strategy, like the fully personalized model, assumes that a calibration procedure will occur for each user of AirWare.}
From  \autoref{fig:bestmodel}, we see that model 3 is the best performing amongst all the models. We are able to achieve an average true positive rate of~50.45\% per class per user with a standard error of~0.03\% across users for the 21-class gesture set. 


 
If we investigate the most common confusions, we  see that \textit{click} gets misclassified as \textit{double-click} 26\%~of the times, \textit{pan down} gets misclassified as \textit{flick down} ~(18\%) and \textit{pan right} gets misclassified as \textit{flick right}~(21\%), suggesting a close similarity between their Doppler signatures and IR activations. The best performing gestures are \textit{erase}, \textit{pans}, and \textit{snap} which have average true positive rates of 89\%, 72\%, and 72\%, respectively.

Finally, we also wish to understand about how much training data is required before the performance of the different models begins to saturate. That is, about how many calibration examples are required before the performance plateaus? To investigate this question we look at the training curves for `Model 3' and `Random Forest' since these are the best performing models.

\begin{figure}[t]
\centering
  \includegraphics[width=0.5\linewidth]{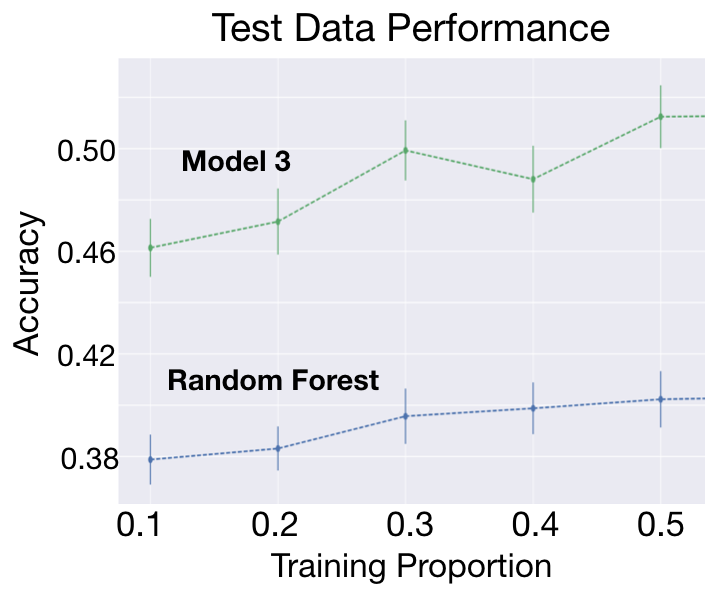}
  \caption{Training Curve for Model 3 and Random Forest using `User-calibrated' cross-validation strategy}
  \label{fig:trainingcurve}
\end{figure}

\autoref{fig:trainingcurve} \hl{shows the performance of `Model 3' and `Random Forest' as we gradually increase the percentage of calibration data from 10\% to 50\%.} \hl{We increase the training size by 10\% and evaluated the models using the remaining data from the user not used in calibration. As we can see in} \autoref{fig:trainingcurve}, \hl{both `Model 3' and the `Random Forest' model gradually increase performance as more user-specific calibration data is added. Moreover, both models begin to saturate between 30\% and 50\% of training data used from the user. If we assume saturation is achieved at 50\%, this corresponds to the system needing 2--3 examples of each gesture from the user during calibration. }

From the above results, we can clearly see that fusing the two different sensing modalities allows increased performance as compared to only using the individual sensor information. However, the overall performance of 50.45\% is still dramatically less than what would be needed for a practical gesture recognition system. We conclude that we cannot support the full 21 gesture vocabulary at a given time. However, it may be possible to select subsets of the gestures from the full vocabulary. Thus, we explore what simplifications to the vocabulary can be made to increase the per gesture true positive rate to a point of usability.

\subsection{Using Feature Subsets}
\begin{figure}[t]
  \includegraphics[width=0.7\linewidth]{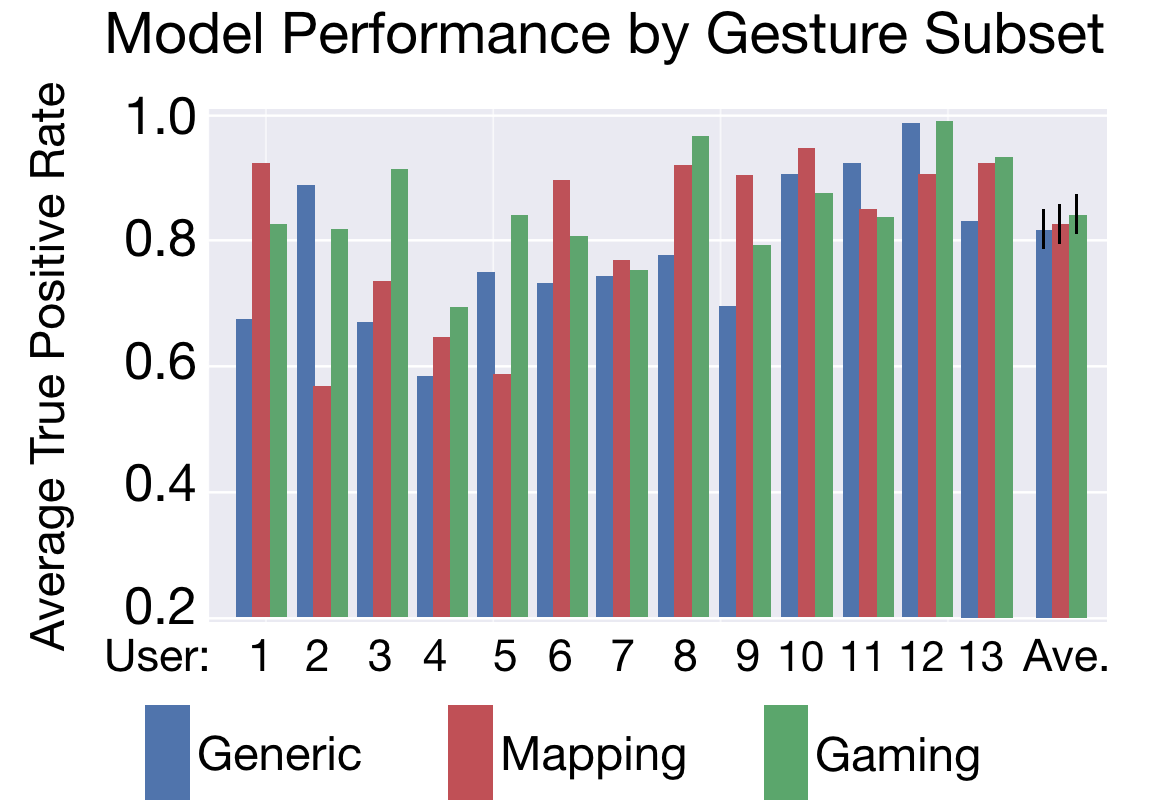}
  \caption{Performance comparison across users for different reduced gesture sets}
  \label{fig:reducedgesture}
\end{figure}

We now seek to understand if the performance of the classifier can be improved by reducing the simultaneous number of gestures that a classifier must distinguish for a given application. In this scenario, we wish to divide the gestures into smaller subsets based upon what combinations of gestures are most appropriate for different categories of applications. We assume that the application in question somehow instructs the user of what gestures are currently supported. If a user were to perform an unsupported gesture, the system would misinterpret that gesture.
Sub-setting the gesture set, we  enumerate 4 different categories: \textit{Generic, Mapping,} and \textit{Gaming}. Each gesture set comprises \hl{4~to 7~gestures}. Together, these  categories include \hl{16~distinct gestures}. Thus, the vocabulary is large, but managed by never having more than \hl{7~gestures} available at a time.
We test the performance of the model for these reduced gesture sets using `user calibrated' strategy discussed above. We also employ the most accurate architecture as selected through previous analysis and parameters remain the same as from our previous hyper-parameter tuning. Confusion matrices are generated in the same manner as previously discussed. \hl{A summary of the different reduced sets overall and per user appears in }\autoref{fig:reducedgesture}. \hl{As shown, there are a number of users for which the system works well for and a number of individuals that it does not always achieve high true positive rate. In particular, users 1, 3, and 4 have reduced recognition rates compared to other users. Upon review of the data, these also corresponded to users that did not perform many practice trials while learning the gestures. These users only practiced the gestures one time compared to other participants performing gestures multiple times before they reported that they were ready to start the experiment. As such, these participants may have rushed through the learning of the gestures or not taken the experiment as seriously as others. }

\begin{figure}[b]
  \includegraphics[width=\linewidth]{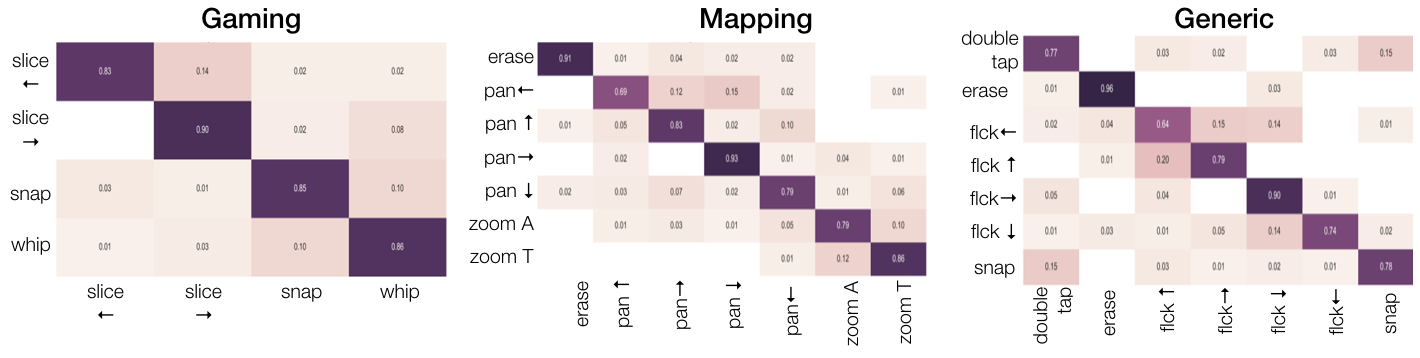}
  \caption{Aggregate confusion matrix for user calibrated performance of the reduced gesture sets.}
  \label{fig:reduced_conf}
\end{figure}

\textbf{\textit{Generic}} is comprised of a total of 7~gestures: \textit{double-tap, flicks up/down/left/right, snap} and \textit{erase}. This set is most likely to be used by interfaces that require generic \hl{up/down/left/right interactions as well as some selection and undo interactions} such as when interacting with a web browser. 
With these 7 gestures, AirWare is able to achieve an average true positive rate of~82.4\% per gesture across users with a standard error of~0.02\%~(\autoref{fig:reducedgesture}). \hl{Inherently, the \textit{double tap} and \textit{snap} gestures are performed in a similar fashion, which accounts for most confusions. From} \autoref{fig:reduced_conf}, \hl{we can see the model confuses between these two gestures often. Similarly, we can see that \textit{flick up} and \textit{flick right} are confused by the model. This is likely the result of these gestures being performed quickly or sloppily by participants---resulting in a number of `diagonal' motions rather than explicitly downward or leftward motions. Because of this, it might be warranted to limit \textit{flick} interactions to only \textit{left/right} or only \textit{up/down}. Furthermore eliminating the \textit{snap} gesture would help increase the true positive rate of the \textit{double tap} gesture.} 


\textbf{\textit{Mapping}} set is focused on more immersive applications such as maps where zooming and panning are core requirements. It consists of \textit{Zoom in/away, Pan up/down/left/right} and \textit{erase} (a total of 7 gestures). 
We are able to achieve an average true positive rate of \hl{82.9\% with a standard error of 0.02.} (See \autoref{fig:reducedgesture}.) From \autoref{fig:reduced_conf}, \hl{we can see the model confuses between \textit{pan down} with \textit{pan left} or \textit{pan right} gestures the most. This is a similar phenomenon to users performing flicks at diagonal angles rather than directly lateral to the phone. Even so, most pans are classified accurately.}

\textbf{\textit{Gaming}} set consists of \textit{snap, slice left/right} and \textit{Whip} (a total of 4 gestures) focusing on specialized applications like gaming. 
We are able to achieve an average true positive rate of \hl{86.5\%} per gesture across users with a \hl{standard error of 0.03\%.} (See \autoref{fig:reducedgesture}.) From \autoref{fig:reduced_conf}, we can see the model confuses between \textit{snap} and \textit{whip} gestures the most, but mostly all gestures are classified accurately.


The reduced gesture sets can be combined in different scenarios and for different applications to achieve a gesture vocabulary of 16 unique in-air gestures with about 80\% or better average true positive rate per gesture.  Many gestures surpass 90\% true positive rate. \hl{The reduced gesture sets presented here are only an example of possible subsets. Depending on the needs of the application, a number of reduced sets could be deployed by the AirWare system for different usage scenarios to support an even larger vocabulary.} 
With this level of performance, we believe AirWare could be used in modern smartphones for a number of application scenarios. However, the full 21 gesture vocabulary is not accurate enough to be deployed by a gesture recognition system. 


\section{Discussions and Future Work}
Despite the good performance of AirWare, there are limitations in our study that we wish to specifically mention. First, our evaluation does not incorporate any specific user interface task. That is, the gesture performed by the participants were in random order and did not have an action task associated to them. It might be possible that when users employ these gestures for specific tasks, the way they are performed may alter in comparison to our user study. Also taking into consideration that the gestures were recorded in a relatively controlled setting, the performance can depreciate in real-life environments having external acoustic disturbances or when the phone is held in different positions. Even so, previous Doppler based research has shown the method to be robust to many acoustic environments~\cite{gupta2012soundwave}.  

We have shown that user specific calibration significantly boosts classification performance. However, this limits the scalability of our approach because it requires users to provide calibration examples. Once calibration examples are collected, the architecture must be retrained. 
This retraining is limited by the current computing power of smart-phones. Considering the complexity of algorithm, calibration would require cloud support to be realistic.  Even so, once trained, the neural network architecture, short-time Fourier transform, and IR  values are computationally efficient and easy to compute from the smart-phone in real-time. The battery considerations for such an implementation are not too much of a concern because gestures only need to be processed when the IR sensor is activated. This helps to further reduce the computational cost of the AirWare approach.  \hl{Even so, because we require the IR sensor to be activated, users may feel like some gestures in the vocabulary are awkward or unintuitive to perform. This limitation would also require that users are instructed on how to perform each of the gestures and then the system would need to verify that the user could complete the gestures reliably. While a limitation, we foresee this process as something that could be incorporated into the calibration phase of the system. } 


We would also like to point out the problems with the Samsung S5 gesture sensing API. Samsung has deprecated support for the device and access to the sensor output is limited. Moreover, there is no way to access the raw sensor values without rooting the phone. This limits the impact of our current approach to pervade the current market, but doesn't limit the research contribution. This deprecation did affect our user study. Because the sensor API was deprecated, many of the angle and velocity measures were flagged ``unknown.'' We removed those incomplete records from our dataset but the reliability of the sensor reading is called into question. 
As such, our results might represent a lower bound of performance and may be further increased with more reliable sensor readings or more expressive IR sensor data. 

 Finally, we have not investigated user adoption of our vocabulary set nor have we investigated impact of our large gesture vocabulary. We leave these limitations to future work.
\section{Conclusions}
In conclusion, we presented AirWare, a technology that fuses the output of an embedded smart-phone microphone and proximity sensor to recognize a gesture set of 21 in-air hand-gestures with 50.47\% average true positive rate per gesture. While we show that combining two different sensor information streams can significantly increase performance, we conclude that the full 21 gesture vocabulary cannot be reliably classified for use in a deployed gesture recognition system. However, AirWare can achieve a reliable true positive rate per gesture for a number of reduced vocabulary gesture sets. In particular, AirWare can achieve true positive rates of greater than 80\% true positive rate for \textit{Generic}, \textit{Mapping}, and \textit{Gaming} gesture sets. 
Using these gesture sets, AirWare can reliably classify a vocabulary of 16 unique gestures, with 4--7 gestures supported at any given time. 

\bibliographystyle{ACM-Reference-Format}
\nocite{*}
\bibliography{citations-bib}


\begin{thebibliography}{31}


\ifx \showCODEN    \undefined \def \showCODEN     #1{\unskip}     \fi
\ifx \showDOI      \undefined \def \showDOI       #1{#1}\fi
\ifx \showISBNx    \undefined \def \showISBNx     #1{\unskip}     \fi
\ifx \showISBNxiii \undefined \def \showISBNxiii  #1{\unskip}     \fi
\ifx \showISSN     \undefined \def \showISSN      #1{\unskip}     \fi
\ifx \showLCCN     \undefined \def \showLCCN      #1{\unskip}     \fi
\ifx \shownote     \undefined \def \shownote      #1{#1}          \fi
\ifx \showarticletitle \undefined \def \showarticletitle #1{#1}   \fi
\ifx \showURL      \undefined \def \showURL       {\relax}        \fi
\providecommand\bibfield[2]{#2}
\providecommand\bibinfo[2]{#2}
\providecommand\natexlab[1]{#1}
\providecommand\showeprint[2][]{arXiv:#2}

\bibitem[\protect\citeauthoryear{Abadi, Agarwal, Barham, Brevdo, Chen, Citro,
  Corrado, Davis, Dean, Devin, Ghemawat, Goodfellow, Harp, Irving, Isard, Jia,
  Jozefowicz, Kaiser, Kudlur, Levenberg, Man\'{e}, Monga, Moore, Murray, Olah,
  Schuster, Shlens, Steiner, Sutskever, Talwar, Tucker, Vanhoucke, Vasudevan,
  Vi\'{e}gas, Vinyals, Warden, Wattenberg, Wicke, Yu, and Zheng}{Abadi
  et~al\mbox{.}}{2015}]%
        {tensorflow2015-whitepaper}
\bibfield{author}{\bibinfo{person}{Mart\'{\i}n Abadi}, \bibinfo{person}{Ashish
  Agarwal}, \bibinfo{person}{Paul Barham}, \bibinfo{person}{Eugene Brevdo},
  \bibinfo{person}{Zhifeng Chen}, \bibinfo{person}{Craig Citro},
  \bibinfo{person}{Greg~S. Corrado}, \bibinfo{person}{Andy Davis},
  \bibinfo{person}{Jeffrey Dean}, \bibinfo{person}{Matthieu Devin},
  \bibinfo{person}{Sanjay Ghemawat}, \bibinfo{person}{Ian Goodfellow},
  \bibinfo{person}{Andrew Harp}, \bibinfo{person}{Geoffrey Irving},
  \bibinfo{person}{Michael Isard}, \bibinfo{person}{Yangqing Jia},
  \bibinfo{person}{Rafal Jozefowicz}, \bibinfo{person}{Lukasz Kaiser},
  \bibinfo{person}{Manjunath Kudlur}, \bibinfo{person}{Josh Levenberg},
  \bibinfo{person}{Dan Man\'{e}}, \bibinfo{person}{Rajat Monga},
  \bibinfo{person}{Sherry Moore}, \bibinfo{person}{Derek Murray},
  \bibinfo{person}{Chris Olah}, \bibinfo{person}{Mike Schuster},
  \bibinfo{person}{Jonathon Shlens}, \bibinfo{person}{Benoit Steiner},
  \bibinfo{person}{Ilya Sutskever}, \bibinfo{person}{Kunal Talwar},
  \bibinfo{person}{Paul Tucker}, \bibinfo{person}{Vincent Vanhoucke},
  \bibinfo{person}{Vijay Vasudevan}, \bibinfo{person}{Fernanda Vi\'{e}gas},
  \bibinfo{person}{Oriol Vinyals}, \bibinfo{person}{Pete Warden},
  \bibinfo{person}{Martin Wattenberg}, \bibinfo{person}{Martin Wicke},
  \bibinfo{person}{Yuan Yu}, {and} \bibinfo{person}{Xiaoqiang Zheng}.}
  \bibinfo{year}{2015}\natexlab{}.
\newblock \bibinfo{title}{{TensorFlow}: Large-Scale Machine Learning on
  Heterogeneous Systems}.
\newblock
\newblock
\urldef\tempurl%
\url{http://tensorflow.org/}
\showURL{%
\tempurl}
\newblock
\shownote{Software available from tensorflow.org.}


\bibitem[\protect\citeauthoryear{Aumi, Gupta, Goel, Larson, and Patel}{Aumi
  et~al\mbox{.}}{2013}]%
        {aumi2013doplink}
\bibfield{author}{\bibinfo{person}{Md~Tanvir~Islam Aumi},
  \bibinfo{person}{Sidhant Gupta}, \bibinfo{person}{Mayank Goel},
  \bibinfo{person}{Eric Larson}, {and} \bibinfo{person}{Shwetak Patel}.}
  \bibinfo{year}{2013}\natexlab{}.
\newblock \showarticletitle{DopLink: using the doppler effect for multi-device
  interaction}. In \bibinfo{booktitle}{\emph{Proceedings of the 2013 ACM
  international joint conference on Pervasive and ubiquitous computing}}. ACM,
  \bibinfo{pages}{583--586}.
\newblock


\bibitem[\protect\citeauthoryear{Bannis, Zhang, and Pan}{Bannis
  et~al\mbox{.}}{2014}]%
        {bannis2014adding}
\bibfield{author}{\bibinfo{person}{Adeola Bannis}, \bibinfo{person}{Pei Zhang},
  {and} \bibinfo{person}{Shijia Pan}.} \bibinfo{year}{2014}\natexlab{}.
\newblock \showarticletitle{Adding directional context to gestures using
  doppler effect}. In \bibinfo{booktitle}{\emph{Proceedings of the 2014 ACM
  International Joint Conference on Pervasive and Ubiquitous Computing: Adjunct
  Publication}}. ACM, \bibinfo{pages}{5--8}.
\newblock


\bibitem[\protect\citeauthoryear{Bengio}{Bengio}{2012}]%
        {bengio2012practical}
\bibfield{author}{\bibinfo{person}{Yoshua Bengio}.}
  \bibinfo{year}{2012}\natexlab{}.
\newblock \showarticletitle{Practical recommendations for gradient-based
  training of deep architectures}.
\newblock In \bibinfo{booktitle}{\emph{Neural networks: Tricks of the trade}}.
  \bibinfo{publisher}{Springer}, \bibinfo{pages}{437--478}.
\newblock


\bibitem[\protect\citeauthoryear{Bergstra, Bardenet, Bengio, and
  K{\'e}gl}{Bergstra et~al\mbox{.}}{2011}]%
        {bergstra2011algorithms}
\bibfield{author}{\bibinfo{person}{James~S Bergstra}, \bibinfo{person}{R{\'e}mi
  Bardenet}, \bibinfo{person}{Yoshua Bengio}, {and} \bibinfo{person}{Bal{\'a}zs
  K{\'e}gl}.} \bibinfo{year}{2011}\natexlab{}.
\newblock \showarticletitle{Algorithms for hyper-parameter optimization}. In
  \bibinfo{booktitle}{\emph{Advances in Neural Information Processing
  Systems}}. \bibinfo{pages}{2546--2554}.
\newblock


\bibitem[\protect\citeauthoryear{Butler, Izadi, and Hodges}{Butler
  et~al\mbox{.}}{2008}]%
        {butler2008sidesight}
\bibfield{author}{\bibinfo{person}{Alex Butler}, \bibinfo{person}{Shahram
  Izadi}, {and} \bibinfo{person}{Steve Hodges}.}
  \bibinfo{year}{2008}\natexlab{}.
\newblock \showarticletitle{SideSight: multi-touch interaction around small
  devices}. In \bibinfo{booktitle}{\emph{Proceedings of the 21st annual ACM
  symposium on User interface software and technology}}. ACM,
  \bibinfo{pages}{201--204}.
\newblock


\bibitem[\protect\citeauthoryear{Chen, Ashbrook, Goel, Lee, and Patel}{Chen
  et~al\mbox{.}}{2014}]%
        {chen2014airlink}
\bibfield{author}{\bibinfo{person}{Ke-Yu Chen}, \bibinfo{person}{Daniel
  Ashbrook}, \bibinfo{person}{Mayank Goel}, \bibinfo{person}{Sung-Hyuck Lee},
  {and} \bibinfo{person}{Shwetak Patel}.} \bibinfo{year}{2014}\natexlab{}.
\newblock \showarticletitle{AirLink: sharing files between multiple devices
  using in-air gestures}. In \bibinfo{booktitle}{\emph{Proceedings of the 2014
  ACM International Joint Conference on Pervasive and Ubiquitous Computing}}.
  ACM, \bibinfo{pages}{565--569}.
\newblock


\bibitem[\protect\citeauthoryear{Chen, Lyons, White, and Patel}{Chen
  et~al\mbox{.}}{2013}]%
        {chen2013utrack}
\bibfield{author}{\bibinfo{person}{Ke-Yu Chen}, \bibinfo{person}{Kent Lyons},
  \bibinfo{person}{Sean White}, {and} \bibinfo{person}{Shwetak Patel}.}
  \bibinfo{year}{2013}\natexlab{}.
\newblock \showarticletitle{uTrack: 3D input using two magnetic sensors}. In
  \bibinfo{booktitle}{\emph{Proceedings of the 26th annual ACM symposium on
  User interface software and technology}}. ACM, \bibinfo{pages}{237--244}.
\newblock


\bibitem[\protect\citeauthoryear{Chollet et~al\mbox{.}}{Chollet
  et~al\mbox{.}}{2015}]%
        {chollet2015keras}
\bibfield{author}{\bibinfo{person}{Fran\c{c}ois Chollet} {et~al\mbox{.}}}
  \bibinfo{year}{2015}\natexlab{}.
\newblock \bibinfo{title}{Keras}.
\newblock \bibinfo{howpublished}{\url{https://github.com/fchollet/keras}}.
\newblock


\bibitem[\protect\citeauthoryear{Dumas, Lalanne, and Oviatt}{Dumas
  et~al\mbox{.}}{2009}]%
        {dumas2009multimodal}
\bibfield{author}{\bibinfo{person}{Bruno Dumas}, \bibinfo{person}{Denis
  Lalanne}, {and} \bibinfo{person}{Sharon Oviatt}.}
  \bibinfo{year}{2009}\natexlab{}.
\newblock \showarticletitle{Multimodal interfaces: A survey of principles,
  models and frameworks}.
\newblock \bibinfo{journal}{\emph{Human machine interaction}}
  (\bibinfo{year}{2009}), \bibinfo{pages}{3--26}.
\newblock


\bibitem[\protect\citeauthoryear{Gupta, Morris, Patel, and Tan}{Gupta
  et~al\mbox{.}}{2012}]%
        {gupta2012soundwave}
\bibfield{author}{\bibinfo{person}{Sidhant Gupta}, \bibinfo{person}{Daniel
  Morris}, \bibinfo{person}{Shwetak Patel}, {and} \bibinfo{person}{Desney
  Tan}.} \bibinfo{year}{2012}\natexlab{}.
\newblock \showarticletitle{Soundwave: using the doppler effect to sense
  gestures}. In \bibinfo{booktitle}{\emph{Proceedings of the SIGCHI Conference
  on Human Factors in Computing Systems}}. ACM, \bibinfo{pages}{1911--1914}.
\newblock


\bibitem[\protect\citeauthoryear{Hassani, Van~Dijk, Ludden, and
  Eertink}{Hassani et~al\mbox{.}}{2011}]%
        {hassani2011touch}
\bibfield{author}{\bibinfo{person}{Anouar~Znagui Hassani},
  \bibinfo{person}{Betsy Van~Dijk}, \bibinfo{person}{Geke Ludden}, {and}
  \bibinfo{person}{Henk Eertink}.} \bibinfo{year}{2011}\natexlab{}.
\newblock \showarticletitle{Touch versus in-air hand gestures: evaluating the
  acceptance by seniors of human-robot interaction}. In
  \bibinfo{booktitle}{\emph{International Joint Conference on Ambient
  Intelligence}}. Springer, \bibinfo{pages}{309--313}.
\newblock


\bibitem[\protect\citeauthoryear{Hilliges, Izadi, Wilson, Hodges,
  Garcia-Mendoza, and Butz}{Hilliges et~al\mbox{.}}{2009}]%
        {hilliges2009interactions}
\bibfield{author}{\bibinfo{person}{Otmar Hilliges}, \bibinfo{person}{Shahram
  Izadi}, \bibinfo{person}{Andrew~D Wilson}, \bibinfo{person}{Steve Hodges},
  \bibinfo{person}{Armando Garcia-Mendoza}, {and} \bibinfo{person}{Andreas
  Butz}.} \bibinfo{year}{2009}\natexlab{}.
\newblock \showarticletitle{Interactions in the air: adding further depth to
  interactive tabletops}. In \bibinfo{booktitle}{\emph{Proceedings of the 22nd
  annual ACM symposium on User interface software and technology}}. ACM,
  \bibinfo{pages}{139--148}.
\newblock


\bibitem[\protect\citeauthoryear{Hinckley}{Hinckley}{2003}]%
        {hinckley2003synchronous}
\bibfield{author}{\bibinfo{person}{Ken Hinckley}.}
  \bibinfo{year}{2003}\natexlab{}.
\newblock \showarticletitle{Synchronous gestures for multiple persons and
  computers}. In \bibinfo{booktitle}{\emph{Proceedings of the 16th annual ACM
  symposium on User interface software and technology}}. ACM,
  \bibinfo{pages}{149--158}.
\newblock


\bibitem[\protect\citeauthoryear{Kim and Moon}{Kim and Moon}{2016}]%
        {kim2016human}
\bibfield{author}{\bibinfo{person}{Youngwook Kim} {and} \bibinfo{person}{Taesup
  Moon}.} \bibinfo{year}{2016}\natexlab{}.
\newblock \showarticletitle{Human detection and activity classification based
  on micro-doppler signatures using deep convolutional neural networks}.
\newblock \bibinfo{journal}{\emph{IEEE Geoscience and Remote Sensing Letters}}
  \bibinfo{volume}{13}, \bibinfo{number}{1} (\bibinfo{year}{2016}),
  \bibinfo{pages}{8--12}.
\newblock


\bibitem[\protect\citeauthoryear{Kim and Toomajian}{Kim and Toomajian}{2016}]%
        {kim2016hand}
\bibfield{author}{\bibinfo{person}{Youngwook Kim} {and} \bibinfo{person}{Brian
  Toomajian}.} \bibinfo{year}{2016}\natexlab{}.
\newblock \showarticletitle{Hand Gesture Recognition Using Micro-Doppler
  Signatures With Convolutional Neural Network}.
\newblock \bibinfo{journal}{\emph{IEEE Access}}  \bibinfo{volume}{4}
  (\bibinfo{year}{2016}), \bibinfo{pages}{7125--7130}.
\newblock


\bibitem[\protect\citeauthoryear{Lee}{Lee}{2010}]%
        {lee2010search}
\bibfield{author}{\bibinfo{person}{Johnny~Chung Lee}.}
  \bibinfo{year}{2010}\natexlab{}.
\newblock \showarticletitle{In search of a natural gesture.}
\newblock \bibinfo{journal}{\emph{ACM Crossroads}} \bibinfo{volume}{16},
  \bibinfo{number}{4} (\bibinfo{year}{2010}), \bibinfo{pages}{9--12}.
\newblock


\bibitem[\protect\citeauthoryear{Lien, Gillian, Karagozler, Amihood, Schwesig,
  Olson, Raja, and Poupyrev}{Lien et~al\mbox{.}}{2016}]%
        {Soli}
\bibfield{author}{\bibinfo{person}{Jaime Lien}, \bibinfo{person}{Nicholas
  Gillian}, \bibinfo{person}{M~Emre Karagozler}, \bibinfo{person}{Patrick
  Amihood}, \bibinfo{person}{Carsten Schwesig}, \bibinfo{person}{Erik Olson},
  \bibinfo{person}{Hakim Raja}, {and} \bibinfo{person}{Ivan Poupyrev}.}
  \bibinfo{year}{2016}\natexlab{}.
\newblock \showarticletitle{Soli: Ubiquitous gesture sensing with millimeter
  wave radar}.
\newblock \bibinfo{journal}{\emph{ACM Transactions on Graphics (TOG)}}
  \bibinfo{volume}{35}, \bibinfo{number}{4} (\bibinfo{year}{2016}),
  \bibinfo{pages}{142}.
\newblock


\bibitem[\protect\citeauthoryear{L{\"o}cken, Hesselmann, Pielot, Henze, and
  Boll}{L{\"o}cken et~al\mbox{.}}{2012}]%
        {locken2012user}
\bibfield{author}{\bibinfo{person}{Andreas L{\"o}cken}, \bibinfo{person}{Tobias
  Hesselmann}, \bibinfo{person}{Martin Pielot}, \bibinfo{person}{Niels Henze},
  {and} \bibinfo{person}{Susanne Boll}.} \bibinfo{year}{2012}\natexlab{}.
\newblock \showarticletitle{User-centred process for the definition of
  free-hand gestures applied to controlling music playback}.
\newblock \bibinfo{journal}{\emph{Multimedia systems}} \bibinfo{volume}{18},
  \bibinfo{number}{1} (\bibinfo{year}{2012}), \bibinfo{pages}{15--31}.
\newblock


\bibitem[\protect\citeauthoryear{Mundada}{Mundada}{2017}]%
        {opensourceRaunak}
\bibfield{author}{\bibinfo{person}{Raunak Mundada}.}
  \bibinfo{year}{2017}\natexlab{}.
\newblock \bibinfo{title}{AirWare Open Source Repository}.
\newblock \bibinfo{howpublished}{\url{https://github.com/raunakm90/AirWare}}.
\newblock


\bibitem[\protect\citeauthoryear{Pedregosa, Varoquaux, Gramfort, Michel,
  Thirion, Grisel, Blondel, Prettenhofer, Weiss, Dubourg,
  et~al\mbox{.}}{Pedregosa et~al\mbox{.}}{2011}]%
        {pedregosa2011scikit}
\bibfield{author}{\bibinfo{person}{Fabian Pedregosa}, \bibinfo{person}{Ga{\"e}l
  Varoquaux}, \bibinfo{person}{Alexandre Gramfort}, \bibinfo{person}{Vincent
  Michel}, \bibinfo{person}{Bertrand Thirion}, \bibinfo{person}{Olivier
  Grisel}, \bibinfo{person}{Mathieu Blondel}, \bibinfo{person}{Peter
  Prettenhofer}, \bibinfo{person}{Ron Weiss}, \bibinfo{person}{Vincent
  Dubourg}, {et~al\mbox{.}}} \bibinfo{year}{2011}\natexlab{}.
\newblock \showarticletitle{Scikit-learn: Machine learning in Python}.
\newblock \bibinfo{journal}{\emph{Journal of Machine Learning Research}}
  \bibinfo{volume}{12}, \bibinfo{number}{Oct} (\bibinfo{year}{2011}),
  \bibinfo{pages}{2825--2830}.
\newblock


\bibitem[\protect\citeauthoryear{Raj, Kalgaonkar, Harrison, and Dietz}{Raj
  et~al\mbox{.}}{2012}]%
        {raj2012ultrasonic}
\bibfield{author}{\bibinfo{person}{Bhiksha Raj}, \bibinfo{person}{Kaustubh
  Kalgaonkar}, \bibinfo{person}{Chris Harrison}, {and} \bibinfo{person}{Paul
  Dietz}.} \bibinfo{year}{2012}\natexlab{}.
\newblock \showarticletitle{Ultrasonic Doppler sensing in HCI}.
\newblock \bibinfo{journal}{\emph{IEEE Pervasive Computing}}
  \bibinfo{volume}{11}, \bibinfo{number}{2} (\bibinfo{year}{2012}),
  \bibinfo{pages}{24--29}.
\newblock


\bibitem[\protect\citeauthoryear{Rautaray and Agrawal}{Rautaray and
  Agrawal}{2015}]%
        {rautaray2015vision}
\bibfield{author}{\bibinfo{person}{Siddharth~S Rautaray} {and}
  \bibinfo{person}{Anupam Agrawal}.} \bibinfo{year}{2015}\natexlab{}.
\newblock \showarticletitle{Vision based hand gesture recognition for human
  computer interaction: a survey}.
\newblock \bibinfo{journal}{\emph{Artificial Intelligence Review}}
  \bibinfo{volume}{43}, \bibinfo{number}{1} (\bibinfo{year}{2015}),
  \bibinfo{pages}{1--54}.
\newblock


\bibitem[\protect\citeauthoryear{Rubine}{Rubine}{1991}]%
        {rubine1991automatic}
\bibfield{author}{\bibinfo{person}{Dean~Harris Rubine}.}
  \bibinfo{year}{1991}\natexlab{}.
\newblock \emph{\bibinfo{title}{The automatic recognition of gestures}}.
\newblock \bibinfo{thesistype}{Ph.D. Dissertation}. \bibinfo{school}{University
  of Toronto}.
\newblock


\bibitem[\protect\citeauthoryear{Severyn and Moschitti}{Severyn and
  Moschitti}{2015}]%
        {severyn2015learning}
\bibfield{author}{\bibinfo{person}{Aliaksei Severyn} {and}
  \bibinfo{person}{Alessandro Moschitti}.} \bibinfo{year}{2015}\natexlab{}.
\newblock \showarticletitle{Learning to rank short text pairs with
  convolutional deep neural networks}. In \bibinfo{booktitle}{\emph{Proceedings
  of the 38th International ACM SIGIR Conference on Research and Development in
  Information Retrieval}}. ACM, \bibinfo{pages}{373--382}.
\newblock


\bibitem[\protect\citeauthoryear{Song, S{\"o}r{\"o}s, Pece, Fanello, Izadi,
  Keskin, and Hilliges}{Song et~al\mbox{.}}{2014}]%
        {song2014air}
\bibfield{author}{\bibinfo{person}{Jie Song}, \bibinfo{person}{G{\'a}bor
  S{\"o}r{\"o}s}, \bibinfo{person}{Fabrizio Pece}, \bibinfo{person}{Sean~Ryan
  Fanello}, \bibinfo{person}{Shahram Izadi}, \bibinfo{person}{Cem Keskin},
  {and} \bibinfo{person}{Otmar Hilliges}.} \bibinfo{year}{2014}\natexlab{}.
\newblock \showarticletitle{In-air gestures around unmodified mobile devices}.
  In \bibinfo{booktitle}{\emph{Proceedings of the 27th annual ACM symposium on
  User interface software and technology}}. ACM, \bibinfo{pages}{319--329}.
\newblock


\bibitem[\protect\citeauthoryear{Starner, Auxier, Ashbrook, and Gandy}{Starner
  et~al\mbox{.}}{2000}]%
        {starner2000gesture}
\bibfield{author}{\bibinfo{person}{Thad Starner}, \bibinfo{person}{Jake
  Auxier}, \bibinfo{person}{Daniel Ashbrook}, {and} \bibinfo{person}{Maribeth
  Gandy}.} \bibinfo{year}{2000}\natexlab{}.
\newblock \showarticletitle{The gesture pendant: A self-illuminating, wearable,
  infrared computer vision system for home automation control and medical
  monitoring}. In \bibinfo{booktitle}{\emph{Wearable computers, the fourth
  international symposium on}}. IEEE, \bibinfo{pages}{87--94}.
\newblock


\bibitem[\protect\citeauthoryear{Suarez and Murphy}{Suarez and Murphy}{2012}]%
        {suarez2012hand}
\bibfield{author}{\bibinfo{person}{Jesus Suarez} {and} \bibinfo{person}{Robin~R
  Murphy}.} \bibinfo{year}{2012}\natexlab{}.
\newblock \showarticletitle{Hand gesture recognition with depth images: A
  review}. In \bibinfo{booktitle}{\emph{RO-MAN, 2012 IEEE}}. IEEE,
  \bibinfo{pages}{411--417}.
\newblock


\bibitem[\protect\citeauthoryear{Sun, Purohit, Bose, and Zhang}{Sun
  et~al\mbox{.}}{2013}]%
        {sun2013spartacus}
\bibfield{author}{\bibinfo{person}{Zheng Sun}, \bibinfo{person}{Aveek Purohit},
  \bibinfo{person}{Raja Bose}, {and} \bibinfo{person}{Pei Zhang}.}
  \bibinfo{year}{2013}\natexlab{}.
\newblock \showarticletitle{Spartacus: spatially-aware interaction for mobile
  devices through energy-efficient audio sensing}. In
  \bibinfo{booktitle}{\emph{Proceeding of the 11th annual international
  conference on Mobile systems, applications, and services}}. ACM,
  \bibinfo{pages}{263--276}.
\newblock


\bibitem[\protect\citeauthoryear{Wobbrock}{Wobbrock}{2006}]%
        {wobbrock2006future}
\bibfield{author}{\bibinfo{person}{Jacob~O Wobbrock}.}
  \bibinfo{year}{2006}\natexlab{}.
\newblock \showarticletitle{The future of mobile device research in HCI}. In
  \bibinfo{booktitle}{\emph{CHI 2006 workshop proceedings: what is the next
  generation of human-computer interaction}}. \bibinfo{pages}{131--134}.
\newblock


\bibitem[\protect\citeauthoryear{Zhao, Chen, Aumi, Patel, and Reynolds}{Zhao
  et~al\mbox{.}}{2014}]%
        {zhao2014sideswipe}
\bibfield{author}{\bibinfo{person}{Chen Zhao}, \bibinfo{person}{Ke-Yu Chen},
  \bibinfo{person}{Md~Tanvir~Islam Aumi}, \bibinfo{person}{Shwetak Patel},
  {and} \bibinfo{person}{Matthew~S Reynolds}.} \bibinfo{year}{2014}\natexlab{}.
\newblock \showarticletitle{SideSwipe: detecting in-air gestures around mobile
  devices using actual GSM signal}. In \bibinfo{booktitle}{\emph{Proceedings of
  the 27th annual ACM symposium on User interface software and technology}}.
  ACM, \bibinfo{pages}{527--534}.
\newblock


\end{thebibliography}

\end{document}